\documentclass[twocolumn,showpacs,preprintnumbers,prl,amsmath,amssymb,floatfix]{revtex4}
\usepackage{epsfig}
\begin{document}
%
%

\title{Exact integral equation for the renormalized Fermi surface}
\author{Sascha Ledowski  and Peter Kopietz}
\affiliation{Institut f\"{u}r Theoretische Physik, Universit\"{a}t Frankfurt,
Robert-Mayer-Strasse 8, 60054 Frankfurt, Germany}
\date{August 27, 2002}
\begin{abstract}
The true Fermi surface of a
fermionic many-body system 
can be viewed as a fixed point manifold
of the renormalization group (RG). 
Within the framework of the exact functional RG 
we show that the fixed point condition implies an exact
integral equation for the counterterm which 
is needed for a self-consistent calculation of the 
Fermi surface.
In the simplest approximation, our integral equation
reduces to the self-consistent Hartree-Fock  equation for the
counterterm.

\end{abstract}
\maketitle

In his authoritative book on interacting Fermi systems
Nozi\`{e}res wrote fourty years ago \cite{Nozieres64}:
``In practice, we shall never try to {\it{calculate}} the Fermi surface, which is much too difficult.'' 
What is the reason  for this difficulty?
Formally, the Fermi surface of an  interacting Fermi system
is defined as the set of all wavevectors ${\bf{k}}_F$
satisfying~\cite{Luttinger60}
 \begin{equation}
 \epsilon_{{ \bf{k}}_F } - \mu + \Sigma ( {\bf{k}}_F , i 0 ) = 0
 \; , 
\label{eq:FSdef}
 \end{equation}
where $\epsilon_{\bf{k}}$ is the energy dispersion in the absence of interactions, 
$\mu$ is the chemical potential, and
$\Sigma ( {\bf{k}} , \omega )$ is the exact self-energy of the interacting system \cite{footnotereal}.
For simplicity we assume an infinite and 
spin-rotationally invariant system at zero temperature, so that 
$\Sigma ( {\bf{k}} , \omega )$ is independent of the spin.
Unfortunately,
 the function $\Sigma ( {\bf{k}}_F , i 0 )$ in Eq. (\ref{eq:FSdef}) is not known a priori, 
so that the calculation of the true  Fermi surface requires
the solution of the many-body problem.

\begin{figure}[tb]
\begin{center}
\epsfig{file=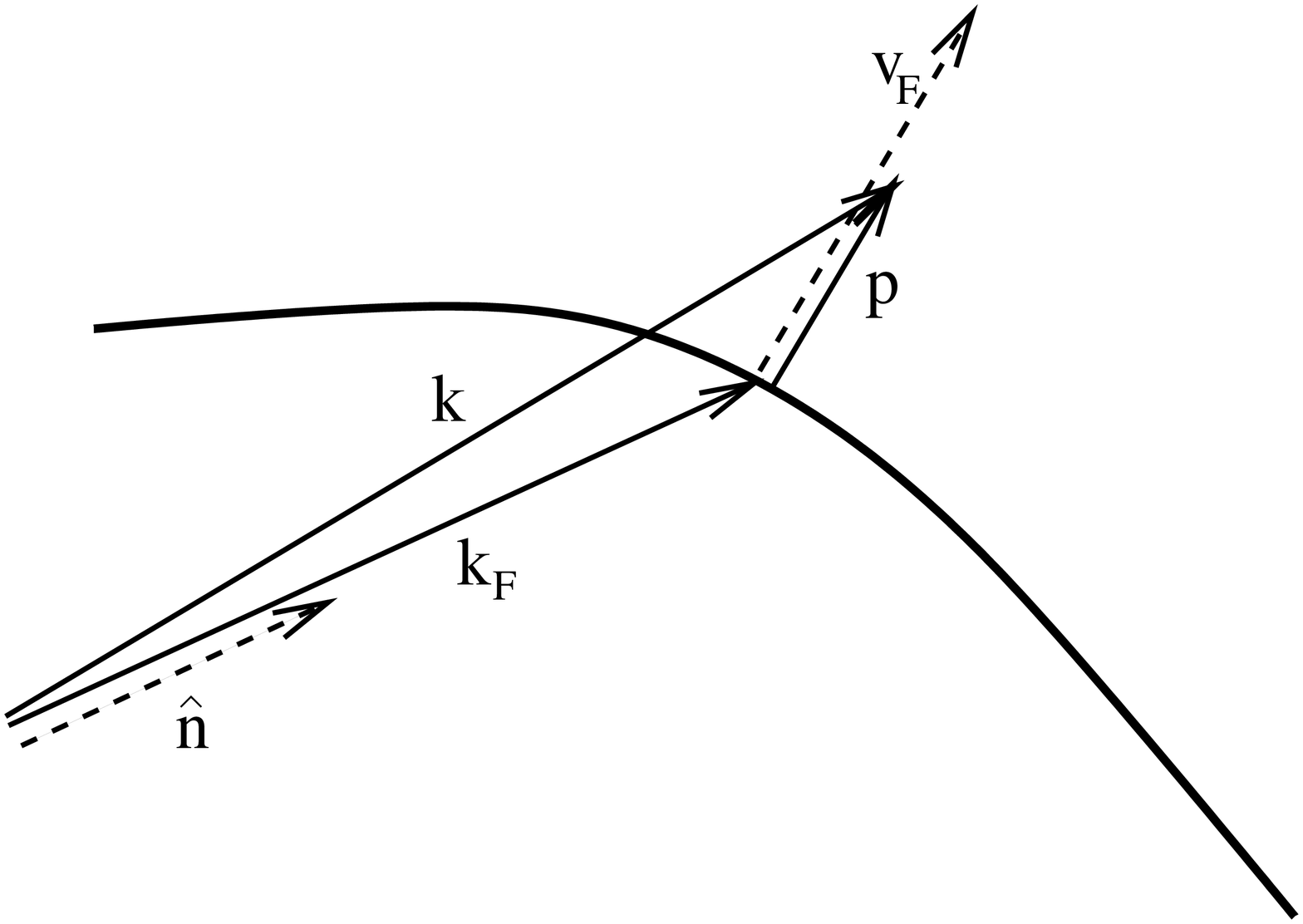,width=70mm}
\end{center}
\vspace{-2mm}
\caption{%
Decomposition ${\bf{k}} = \hat{\bf{n}} k_F ( \hat{\bf{n}} ) +  {\bf{p}} $
of a wavevector ${\bf{k}}$ into a component $ {\bf{k}}_F = \hat{\bf{n}} k_F ( \hat{\bf{n}} )$ on the Fermi surface
and a component $ {\bf{p}} $ in the direction of the 
local Fermi velocity ${\bf{v}}_F$.
The thick solid line is a part of the Fermi surface. This construction
defines ${\bf{k}}_F$ and $\hat{\bf{n}}$ as a function of ${\bf{k}}$.
Here ${\bf{v}}_F = \nabla_{ {\bf{k}}} \epsilon_ { {\bf{k}} }  |_{   {\bf{k}} =  {\bf{k}}_F } $ 
is defined in terms of the gradient of the free dispersion at the true Fermi surface, so that
${\bf{v}}_F$ is not necessarily perpendicular to the Fermi surface.
}
\label{fig:Fermisurface}
\end{figure}

For weak interactions, one might try to determine
the Fermi surface
perturbatively by simply calculating
$ \Sigma ( {\bf{k}} , i 0 )$ in powers of the interaction and
substituting the result into Eq. (\ref{eq:FSdef}).
However, in general the perturbation series
contains anomalous terms \cite{Kohn60} with
unphysical singularities,
which are generated because
the {\it{ground state}} of the non-interacting
system evolves into an {\it{excited state}} of the interacting system
when the interaction is adiabatically switched on.
As discussed by Nozi\`{e}res \cite{Nozieres64},
this artificial level crossing can be avoided by introducing
counterterms which are determined by
the   requirement that
the Fermi surface remains fixed as the interaction is adiabatically switched on. 
This intuitive idea  
can be implemented perturbatively as follows \cite{Nozieres64,Feldman96}:
Suppose we would like to know  the true Fermi surface of a system with 
Hamiltonian
${H} = {H}_0 + {H}_1$, where ${H}_1$ describes some general two-body interaction and
the non-interacting part is given by
 \begin{equation}
 {H}_0   
 = \sum_{ {\bf{k}} ,  \sigma}  \epsilon_{\bf{k}}  c^{\dagger}_{ {\bf{k}} \sigma }
 c_{ {\bf{k}} \sigma } 
 \; .
 \label{eq:H0def}
 \end{equation}
Here $c_{ {\bf{k}} \sigma }$ are the usual annihilation operators of  
fermions with momentum ${\bf{k}}$ and spin $\sigma$. 
An expansion in powers of $H_1$  leads to Feynman diagrams
where vertices corresponding to $H_1$ are connected by 
propagators 
$G_0 ( {\bf{k}} ,  \omega ) = [  \omega - \epsilon_{\bf{k}} + \mu ]^{-1}$.
These  are singular for $\omega = 0$ and
 $ \epsilon_{\bf{k}} = \mu$, which is {\it{not}}
the true Fermi surface
defined in Eq. (\ref{eq:FSdef}).
If the perturbative expansion is truncated at a finite order, this leads to
the unphysical divergencies mentioned above \cite{Kohn60}. To avoid these, we
add the counterterm
$\sum_{ {\bf{k}}  \sigma} \Sigma ( {\bf{k}}_F , i 0 ) 
c^{\dagger}_{ {\bf{k}} \sigma }
 c_{ {\bf{k}} \sigma }  $ to $H_0$ and subtract it again from $H_1$, writing 
 $H = H_0^{\prime} + H_1^{\prime}$, with
 \begin{eqnarray}
 {H}_0^{\prime}  
  =  \sum_{ {\bf{k}}   \sigma}  [ \epsilon_{\bf{k}}  + 
 \Sigma ( {\bf{k}}_F , i 0 )  ]
c^{\dagger}_{ {\bf{k}} \sigma }
 c_{ {\bf{k}} \sigma } 
  \; ,
 \label{eq:H0primedef}
 \end{eqnarray}
and $ H_1^{\prime} = H_1 -
 \sum_{ {\bf{k}} \sigma} \Sigma ( {\bf{k}}_F , i 0 ) 
c^{\dagger}_{ {\bf{k}} \sigma }
 c_{ {\bf{k}} \sigma }  $.
Here ${\bf{k}}_F$ is the wavevector closest to ${\bf{k}}$
lying on the Fermi surface, see Fig. \ref{fig:Fermisurface}.
Using Eq. (\ref{eq:FSdef}),  the
corresponding free propagator can then be written as 
$G_0^{\prime} ( {\bf{k}} ,  \omega ) = [ \omega - \epsilon_{\bf{k}} +
 \epsilon_{  {\bf{k}}_F } ]^{-1}$, which  by construction is  singular on the true Fermi surface.
But how do we find the counterterm necessary to calculate  $\epsilon_{ {\bf{k}}_F }$?
Following the usual strategy adopted in field theory \cite{ZinnJustin89}, 
we may expand the irreducible self-energy associated with 
the modified interaction $H_1^{\prime}$ perturbatively in powers of  $H_1^{\prime}$
and require that, order by order in perturbation theory, 
the corrections vanish when we set $\omega = 0$ and 
${\bf{k}} = {\bf{k}}_F$. This renormalized
perturbation theory
leads to complicated integral equations for
$\Sigma ( {\bf{k}}_F , i0 )$, which must be solved 
numerically \cite{Neumayr02}. 

A few years ago Anderson \cite{Anderson93}
critically discussed 
a simplified version
of the  renormalized perturbation theory outlined above, 
where  $\Sigma ( {\bf{k}}_F , i0 )$ is replaced by a constant $\delta \mu $.
He correctly pointed out that in general it is not allowed
to ignore the momentum-dependence of the counterterms, and
argued that in two dimensions
the effective interaction between 
electrons with opposite spin and wavevectors on the true Fermi surface depends
in a subtle way on the boundary conditions
 which cannot be adequately
taken into account perturbatively.
To gain a better understanding of this problem, it should be useful
to have  an  algorithm for calculating  the Fermi surface
which does not rely on the
perturbative expansion of counterterms in powers of the interaction. 
In this work we show that
such an algorithm follows in a straightforward way from
the  exact functional renormalization group (RG) approach 
described  in Ref.\cite{Kopietz01}.

In the past decade several
authors have used
 Wilsonian RG methods to study 
interacting Fermi systems 
\cite{Kopietz01,Benfatto90,Polchinski92,Shankar94,Zanchi96,Salmhofer98,Halboth00,Binz02},
using different versions of the RG. In particular, 
the RG transformations used in Refs. \cite{Zanchi96,Salmhofer98,Halboth00,Binz02}
exclusively focus on the mode elimination step.
Although such a procedure is sufficient
if one considers only the one-loop flow of marginal couplings,
in general a complete Wilsonian RG transformation consists not only of the 
mode elimination, but 
includes also  the rescaling  of momenta, frequencies and fields \cite{Ma76}.
While the field rescaling (i.e. the wavefunction renormalization)
becomes only important
beyond the one-loop approximation,
the RG flow of all relevant and irrelevant couplings
is determined  in an essential way  by the rescaling step
even at the one-loop level.
We emphasize that the rescaling of momenta and frequencies
is more than a trivial mathematical change of variables --
it is crucial to detect possible fixed points of the  RG and to calculate
critical exponents \cite{Ma76}. 
Moreover, as will be shown below, the length and time rescaling is
very helpful for a derivation of the
self-consistent Hartree-Fock approximation 
within the framework of the exact RG.
The importance of rescaling in the presence of a Fermi surface has 
been emphasized by Polchinski \cite{Polchinski92} and by Shankar \cite{Shankar94}.
In Ref.\cite{Kopietz01} we have shown how 
the exact functional RG approach developed previously \cite{Zanchi96,Salmhofer98,Halboth00}
can be modified to include the rescaling step.
Here, we shall show that the rescaling is crucial
to obtain a non-perturbative definition 
of the Fermi surface as a fixed point manifold
of the RG.  
We obtain
an explicit algorithm for calculating the Fermi surface which does not rely on the iterative
procedure of fixing the counterterms order by order in perturbation theory. 
The fixed point property of the Fermi surface has also
been emphasized by Ferraz \cite{Ferraz02}, who recently discussed
the Fermi surface renormalization in a
special two-dimensional system using the
field theoretical RG.

The exact functional RG can be formulated in terms of an   
infinite hierarchy of  coupled differential equations for the irreducible $2n$-point vertices
$\Gamma^{(2n)}_{\xi} ( K_1^{\prime} , \ldots , K_n^{\prime} ; K_n , \ldots , K_1 )$, where
$K = ( \sigma , {\bf{k}} , i \omega  )$ is a collective label for  spin projection $\sigma$,
momentum ${\bf{k}}$, Matsubara frequency $i \omega $.
Here $\xi$ is an infrared cutoff with units of energy which regularizes the
singularity of the free propagator.
To scale wavevectors toward  the Fermi surface, it is convenient
to perform a
non-linear coordinate transformation in momentum space,
${\bf{k}} = \hat{\bf{n}} k_F ( \hat{\bf{n}} ) + \hat{\bf{v}}_F \xi q / | {\bf{v}}_F |$,
and eliminate ${\bf{k}}$ in favor of the dimensionless variable $q$ and the  unit vector
 $\hat{\bf{n}}$.
Here 
$ k_F ( \hat{\bf{n}} )$ is the length of ${\bf{k}}_F$ parameterized by $\hat{\bf{n}} $,
and ${\bf{v}}_F = \nabla_{\bf{k}} \epsilon_{\bf{k}} |_{ {\bf{k}}_F}$ 
is the Fermi velocity of the non-interacting
system at the true ${\bf{k}}_F$,
see Fig.  \ref{fig:Fermisurface}.
The corresponding unit vector is denoted by
$\hat{\bf{v}}_F = {\bf{v}}_F / | {\bf{v}}_F |$.
Geometrically, $q = {\bf{v}}_F \cdot  {\bf{p}} / \xi =
{\bf{v}}_F \cdot ( {\bf{k}} - \hat{\bf{n}}  k_F  ( \hat{\bf{n}} ) )/ \xi$  measures the 
distance of a given ${\bf{k}}$ from the Fermi surface in units of the infrared cutoff.
We also define rescaled frequencies $\epsilon = \omega / \xi$, and label 
the degrees of freedom by  $Q = ( \sigma ,  \hat{\bf{n}} ,  q , i \epsilon  )$ instead of
$K$. To implement the {\it{scaling toward the Fermi surface}} \cite{Polchinski92}
we consider the RG flow of the following  rescaled vertices \cite{Kopietz01}
\begin{eqnarray}
\tilde{\Gamma}_{t}^{ (2n) }
( Q_1^{\prime} , \ldots , Q_n^{\prime} ; Q_n , \ldots , Q_1 ) =
& & 
 \nonumber \\
& & \hspace{-40mm} \nu_0^{n-1} \xi^{n-2}
\left[ Z_t^{\hat{\bf{n}}_{1}^{\prime} } 
\cdots Z_t^{\hat{\bf{n}}_{n}^{\prime}}
 Z_t^{\hat{\bf{n}}_{n} } \cdots Z_t^{\hat{\bf{n}}_{1}} 
 \right]^{1/2}
 \nonumber
 \\
 & & \hspace{-40mm} \times
\Gamma^{(2n)}_{\xi} ( K_1^{\prime} , \ldots , K_n^{\prime} 
; K_n , \ldots ,
K_1 )
 \; .
 \label{eq:Gammarescale}
\end{eqnarray}
Here $t = - \ln ( \xi / \xi_0 )$ is a logarithmic flow parameter, 
$\nu_0 = \int \frac{ d {\bf{k}}}{ (2 \pi )^D} \delta ( 
\epsilon_{  {\bf{k}} } - \epsilon_{  {\bf{k}}_F }   )  $ is the
density of states (per spin projection) of the non-interacting
system at the true Fermi surface, and
\begin{equation}
Z_t^{\hat{\bf{n}}} = \left[ 1 - \left. \frac{\partial 
\Sigma_{\xi} ( {\bf{k}}_F ,  \omega + i 0)}{\partial
 \omega  } \right|_{\omega=0} \right]^{-1}
\label{eq:Zdef}
\end{equation}
is the wavefunction renormalization factor, where
$\Sigma_{\xi} ( {\bf{k}} , i \omega )$ is the irreducible self-energy of the system with
infrared cutoff $\xi = \xi_0 e^{-t}$. To determine the true
Fermi surface self-consistently it turns out to be useful to  subtract the counterterm
$\Sigma ( {\bf{k}}_F , i0 )$
from the irreducible two-point vertex\cite{Kopietz01}, defining
 \begin{equation}
  \tilde{\Gamma}_{t}^{(2)}
 ( Q  )  = \frac{Z_t^{\hat{\bf{n}} } }{\xi} \Gamma^{(2)}_{\xi} ( K ) =
 - \frac{Z_t^{\hat{\bf{n}} } }{\xi} 
 \left[ \Sigma_{\xi} ( K ) - \Sigma ( {\bf{k}}_F , i0 ) \right]
 \; .
 \label{eq:Gammatdef}
 \end{equation}
The subtracted two-point vertex satisfies the exact flow equation \cite{Kopietz01}
 \begin{eqnarray}
 \partial_t \tilde{\Gamma}_t^{(2)} ( Q ) & = &
 ( 1 - \eta_t^{\hat{\bf{n}}} -    q \partial_q - \epsilon \partial_{\epsilon}    )
 \tilde{\Gamma}_t^{(2)} ( Q )  
 \nonumber
 \\
 & - &  \int_{Q^{\prime}}
 \dot{G}_t ( Q^{\prime} ) 
 \tilde{\Gamma}_{t}^{(4)}
 ( Q , Q^{\prime} ; Q^{\prime} , Q )
 \; ,
 \label{eq:twopointscale}
 \end{eqnarray}
where
$
\eta_t^{\hat{\bf{n}}} = - \partial_t \ln Z_t^{\hat{\bf{n}}}$
is the flowing anomalous dimension (which vanishes for large $t$ if the system is a Fermi liquid).
The integration measure in Eq. (\ref{eq:twopointscale}) is defined by
\begin{equation}
  \int_Q
=  \sum_{\sigma}
\int \frac{ d S_{\hat{\bf{n}} }}{S_D}
\int dq J ({\hat{\bf{n}}} , q ) \int \frac{ d \epsilon}{2 \pi}
\label{eq:measure}
\; \; ,
\end{equation}
where $dS_{\hat{\bf{n}}}$ is a surface element and $S_D$
is the surface area of
the unit sphere in $D$ dimensions, and
 $J ( { \hat{\bf{n}} } ,  q )$ is the Jacobian
associated with the transformation ${\bf{k}} \rightarrow 
( \hat{\bf{n}} , q )$ divided by  $\nu_0 \xi^2$.
The function $ \dot{{G}}_t  ( Q )$ in Eq. (\ref{eq:twopointscale}) is defined by
 \begin{equation}
 \dot{{G}}_t  ( Q ) = \frac{ \delta ( \tilde{\Omega}_Q   -1 )}{
     Z_t^{\hat{\bf{n}}}  \left[ i \epsilon - 
 \xi_t^{ \hat{\bf{n}}} ( q ) \right] + 
 \tilde{\Gamma}_t^{(2)} ( Q )    }
 \; ,
 \end{equation}
where $ 
\xi_t^{ \hat{\bf{n}} } ( q  ) = 
( \epsilon_{ {\bf{k}}_t  } - 
\epsilon_{ {\bf{k}}_F } ) / {\xi}$, with 
${\bf{k}}_t = \hat{\bf{n}} k_F ( \hat{\bf{n}} ) + \hat{\bf{v}}_F \xi_0 e^{-t} q / | {\bf{v}}_F |$.
Here 
$\tilde{\Omega}_Q$ is some function that measures the distance from the Fermi surface, for example
$\tilde{\Omega}_Q = |\xi_t^{ \hat{\bf{n}} } ( q  ) | \approx |q |$.
The  flow equation of the rescaled irreducible four-point vertex
 $\tilde{\Gamma}_{t}^{ (4)}
 ( Q_1^{\prime} , Q_2^{\prime} ; Q_2 , Q_1 ) $
is not  explicitly needed in this work; it can be found in
Ref.\cite{Kopietz01}. 
 
The shape of the Fermi surface of the interacting system 
is determined by the RG flow of the couplings
 \begin{equation}
 \tilde{\mu}_t^{\hat{\bf{n}}}  \equiv  \tilde{\Gamma}_{t}^{(2)}
 ( \sigma , \hat{\bf{n}}  ,   q= 0, i \epsilon = i0   )
 \label{eq:mudef}
 \; .
 \end{equation}
From Eq. (\ref{eq:twopointscale}) we see
that these couplings satisfy the exact flow equation
 \begin{equation}
 \partial_t \tilde{\mu}_t^{\hat{\bf{n}}} =  ( 1 - \eta_t^{\hat{\bf{n}}} ) \tilde{\mu}_t^{\hat{\bf{n}}}
 + \dot{\Gamma}_t^{(2)} ( \hat{\bf{n}} )
 \; ,
 \label{eq:muexact}
 \end{equation}
with
 \begin{equation}
 \dot{\Gamma}_t^{(2)} ( \hat{\bf{n}} ) =
 - \int_{Q^{\prime}}
 \dot{G}_t ( Q^{\prime} ) 
 \tilde{\Gamma}_{t}^{(4)}
 ( Q_0   , Q^{\prime} ; Q^{\prime} ,  Q_0 )
 \; ,
 \end{equation}
where $Q_0 = (  \sigma ,  \hat{\bf{n}} , q=0, i \epsilon = i0  )$.
Note that in dimensions $D > 1$ 
there  are infinitely many couplings $\tilde{\mu}_t^{\hat{\bf{n}}}$,  labelled by
the unit vector $\hat{\bf{n}}$.
Obviously, each $\tilde{\mu}_t^{\hat{\bf{n}}}$ with
$ \eta_{\infty}^{ \hat{\bf{n}}} =
 \lim_{ t \rightarrow \infty } \eta_t^{\hat{\bf{n}}}  < 1 $
is relevant, so that
some fine tuning of the bare couplings $\tilde{\mu}_0^{\hat{\bf{n}}}$ 
is necessary to force $\tilde{\mu}_t^{\hat{\bf{n}}}$    to flow into a fixed point of the RG.
Because a finite limit  $\tilde{\mu}_\infty^{\hat{\bf{n}}} = \lim_{ t \rightarrow \infty  }
\tilde{\mu}_t^{\hat{\bf{n}}} $ means that we have found the true  Fermi surface of the 
interacting system \cite{Kopietz01},
we conclude that the detailed shape of the Fermi surface 
is sensitive to the numerical values of the bare couplings of the theory.
Polchinski  \cite{Polchinski92}  pointed out that such a fine tuning of  the bare couplings
is {\it{unnatural}}, because 
physical effects which  depend on the
precise shape of the Fermi surface 
are not protected against small perturbations. 

To further elucidate the relation between the 
relevant couplings $\tilde{\mu}_t^{\hat{\bf{n}}}$  and the shape of the Fermi surface,
let us explicitly derive from Eq. (\ref{eq:muexact})  
a self-consistency condition for the Fermi surface.   
Therefore it is useful to transform  Eq. (\ref{eq:muexact})  into an integral equation,
 \begin{equation}
 \tilde{\mu}_t^{\hat{\bf{n}} } = e^{ t - \int_0^t d \tau \eta_{\tau}^{\hat{\bf{n}}} }
 \left[
  \tilde{\mu}_0^{\hat{\bf{n}} } +
 \int_{0}^{t} d t^{\prime} e^{ - t^{\prime}   + \int_0^{t^{\prime}} d \tau 
 \eta_{\tau}^{\hat{\bf{n}}} }
 \dot{\Gamma}^{(2)}_{t^{\prime}} ( \hat{\bf{n}} )
 \right]
 \; .
 \label{eq:integral}
 \end{equation}
Suppose now that we have adjusted the bare couplings such that 
for $t \rightarrow \infty $
the flowing couplings $\tilde{\mu}_t^{\hat{\bf{n}} }$ indeed approach finite fixed point values.  
Assuming that the associated anomalous dimensions
$ \eta_{\infty}^{\hat{\bf{n}}}  $
are smaller than unity\cite{footnoteeta}, 
we conclude from Eq. (\ref{eq:integral}) that 
the limit $ \tilde{\mu}_\infty^{\hat{\bf{n}} } =  \lim_{ t \rightarrow \infty } 
\tilde{\mu}_t^{\hat{\bf{n}} } $ can only be finite if the initial values 
$\tilde{\mu}_0^{ \hat{\bf{n}} }$ are chosen such that
 \begin{eqnarray} 
 \tilde{\mu}_0^{\hat{\bf{n}} } & = & -
 \int_{0}^{\infty} d t  e^{ - t   + \int_0^{t} d \tau 
 \eta_{\tau}^{\hat{\bf{n}}} }
 \dot{\Gamma}^{(2)}_{t} ( \hat{\bf{n}} )
 \nonumber
 \\
 &  = & \int_{0}^{\infty} d t  e^{ - t   + \int_0^{t} d \tau 
 \eta_{\tau}^{\hat{\bf{n}}} } 
  \int_{Q^{\prime}}
 \dot{G}_t ( Q^{\prime} ) 
 \tilde{\Gamma}_{t}^{(4)}
 ( Q_0   , Q^{\prime} ; Q^{\prime} ,  Q_0 )
 \; .
 \nonumber
 \\
 & &
 \label{eq:integral2}
 \end{eqnarray}
This is an implicit equation for $ \tilde{\mu}_0^{\hat{\bf{n}} } $, relating
it to the values of the two-point vertex and the four-point vertex
on the entire RG trajectory. Keeping in mind that
the right-hand side of Eq. (\ref{eq:integral2}) 
implicitly depends on $\tilde{\mu}^{ {\hat{\bf{n}} }}_t$ and that
according to Eq. (\ref{eq:Gammatdef})
 \begin{equation}
   \Sigma ( {\bf{k}}_F , i0 )  -     \Sigma_{\xi_0} ( {\bf{k}}_F , i0 )  =
  \xi_0 \tilde{\mu}_0^{\hat{\bf{n}}} /  Z_0^{\hat{\bf{n}} } 
 \; ,
 \label{eq:mugamma}
 \end{equation}
it is obvious  that  Eq. (\ref{eq:integral2}) 
can be regarded as an integral equation for the
counterterm  $ \Sigma ( {\bf{k}}_F , i0 )$, the solution of which yields
the true shape of the  Fermi surface.
At this point  
it is instructive to transform Eq. (\ref{eq:integral2}) back to unrescaled variables,
choosing for simplicity the initial conditions
$ \Sigma_{\xi_0} ( {\bf{k}}_F , i0 ) = 0$ and $ Z_0^{\hat{\bf{n}}} = 1$.
Using the above definitions we find that Eq. (\ref{eq:integral2}) is
equivalent with
 \begin{eqnarray}
 \Sigma ( {\bf{k}}_F , i0 ) & = &  \sum_{\sigma^{\prime} } 
 \int \frac{ d {\bf{k}}^{\prime}   }{ (2 \pi )^{D}} 
 \frac{ d \omega^{\prime} }{ 2 \pi }
 \frac{ \Theta ( \xi_0 - \xi_{ {\bf{k}}^{\prime} } ) }{ i \omega^{\prime} - 
 \epsilon_{ {\bf{k}}^{\prime}} 
 + \mu - \Sigma_{ \xi_{ {\bf{k}}^{\prime}   } } ( K^{\prime} ) }
 \nonumber
 \\
 & & \hspace{-17mm} \times
 \Gamma^{(4)}_{  \xi_{ {\bf{k}}^{\prime}  } }  ( {\bf{k}}_F , i0 , \sigma  , 
 {\bf{k}}^{\prime}, i \omega^{\prime} , \sigma^{\prime}  ; 
   {\bf{k}}^{\prime}, i \omega^{\prime} , \sigma^{\prime} ,  {\bf{k}}_F , i0 , \sigma  )
 \; ,
 \label{eq:counterint}
 \end{eqnarray}
where $  \xi_{ {\bf{k}}^{\prime}  } = | \epsilon_{ {\bf{k}}^{\prime} }  - 
  \epsilon_{ {\bf{k}}_F^{\prime}} | $.
Note that the right-hand side of Eq. (\ref{eq:counterint}) involves the flowing self-energy and
four-point vertex at the scales $ \xi = \xi_{ {\bf{k}}^{\prime} }$ which depend on the
distance from the true Fermi surface.
We emphasize that the exact integral equation 
(\ref{eq:counterint}) and the equivalent rescaled 
equation (\ref{eq:integral2}) 
fix the counterterm  $\Sigma ( {\bf{k}}_F , i0 )$ from the
requirement that for $ t \rightarrow \infty$ 
all couplings approach finite fixed point values.

If the system is a Luttinger liquid, then the rescaled version (\ref{eq:integral2}) of the
self-consistency equation is more convenient than the unrescaled version (\ref{eq:counterint}), 
because for a Luttinger liquid the marginal part of the four-point vertex vertex
$\Gamma^{(4)}_{\xi}$ without wavefunction renormalization factors
diverges for $\xi \rightarrow 0$, while the rescaled four-point
vertex $\tilde{\Gamma}^{(4)}_t$ defined in
Eq.(\ref{eq:Gammarescale})
approaches for  $t \rightarrow \infty$ a finite limit.
In this case the divergence of
the unrescaled $\Gamma^{(4)}_{\xi}$
is canceled by the vanishing wavefunction
renormalization factors at the Luttinger liquid fixed point \cite{Busche02}.

Given a solution 
$ \Sigma ( {\bf{k}}_F , i0 )$
of Eq. (\ref{eq:counterint}), we may calculate the compressibility
of the system by substituting
the result for $ \Sigma ( {\bf{k}}_F , i0 )$ into 
Eq. (\ref{eq:FSdef}) and solving for
$k_F ( \hat{\bf{n}} )$, which implicitly depends
on the chemical potential $\mu$. 
According to the Luttinger theorem \cite{Luttinger60} 
the density $n ( \mu )$ of the system is determined by 
the volume enclosed by the Fermi surface,
 \begin{equation}
  n ( \mu ) = \sum_{\sigma } 
 \int \frac{ d {\bf{k}}   }{ (2 \pi )^{D}} \Theta ( k_F ( \hat{\bf{n}}  ) - | {\bf{k}} | )
 \; ,
 \label{eq:FSdens}
 \end{equation}
 so that we obtain for the compressibility $\chi_n$
 \begin{equation}
 n^2 \chi_n =
  \frac{ \partial n}{\partial \mu } =     \sum_{\sigma}   
\int \frac{ d S_{ {\hat{\bf{n}}} } }{ (2 \pi )^D}
 k_F^{D-1} ( \hat{\bf{n}} ) \frac{ \partial k_F ( \hat{\bf{n}} ) }{\partial \mu }
 \; .
 \label{eq:compres}
 \end{equation}
Note that the compressibility is a 
functional of the shape of the true Fermi surface
of the  many-body system, i.e. the compressibility characterizes the RG fixed point. 
Because the rescaling of  momenta and frequencies is crucial to 
obtain  fixed points of the RG \cite{Ma76},
the compressibility and other uniform susceptibilities
are not so easy to calculate within alternative
versions of the RG which 
omit the rescaling \cite{Salmhofer98,Halboth00,Binz02}.

From our point of view, the Fermi surface is a {\it{fixed point}} manifold of the RG, 
so that it is meaningless to talk about the RG flow of the Fermi surface.
However, it is possible to define a ``flowing Fermi surface'' ${\bf{k}}_{F,t}$  via \cite{Kopietz01}
\begin{equation}
 \epsilon_{{ \bf{k}}_{F,t} } - \mu + \Sigma_{ \xi_0 e^{-t}} ( {\bf{k}}_{F,t} , i 0 ) = 0
 \label{eq:FSdefaux}
 \; ,
 \end{equation} 
which by construction approaches the true Fermi surface 
for $t \rightarrow \infty$.
If we choose the initial conditions at $t=0$ (corresponding to $\xi = \xi_0$)  such that 
$\Sigma_{ \xi_0 } ( {\bf{k}}_{F,0} , i 0 ) = 0$, then 
${\bf{k}}_{F,0}$ is  the Fermi surface of the non-interacting system at the
{\it{same chemical potential}}
as the interacting system.  
This corresponds to a non-interacting system at the 
density
$n_0 ( \mu ) = \sum_{\sigma } 
 \int \frac{ d {\bf{k}}   }{ (2 \pi )^{D}} \Theta ( k_{F,0} ( \hat{\bf{n}}  ) - | {\bf{k}} | )$,
which is in general different from
the density $n ( \mu )$ 
of the interacting system given in Eq. (\ref{eq:FSdens}).
Note that
in Fermi liquid theory one usually works at constant density.
However, as discussed by  Nozi\`{e}res \cite{Nozieres237}, 
for a self-consistent calculation of the Fermi surface it is more convenient
to determine the counterterm at constant chemical potential $\mu$, and
calculate the corresponding density afterwards \cite{footnotedens}.
The practical advantages of such a procedure have  been recognized previously in
Ref.  \cite{Gonzalez00}. Moreover, in the field-theoretical approach 
advanced by Ferraz \cite{Ferraz02} the chemical potential is also a RG invariant.

In the simplest approximation
 Eqs. (\ref{eq:integral2}) and (\ref{eq:counterint}) reduce to
the Hartree-Fock  self-consistency equation for the counterterm
$ \Sigma ( {\bf{k}}_F , i0 )$.
To see this, we approximate  the flowing four-point vertex
in Eq. (\ref{eq:counterint}) as follows 
 \begin{eqnarray}
 \Gamma^{(4)}_{  \xi_{ {\bf{k}}^{\prime}  } }  ( {\bf{k}}_F , i0 , \sigma  , 
 {\bf{k}}^{\prime}, i \omega^{\prime} , \sigma^{\prime}  ; 
   {\bf{k}}^{\prime}, i \omega^{\prime} , \sigma^{\prime} , {\bf{k}}_F , i0 , \sigma  )
  &  &
 \nonumber
 \\ 
 & & \hspace{-72mm} \approx \Gamma^{(4)}_{  \xi = 0  }  ( {\bf{k}}_F , i0 , \sigma  , 
 {\bf{k}}^{\prime}_F , i0  , \sigma^{\prime}  ; 
   {\bf{k}}^{\prime}_F , i 0 , \sigma^{\prime} , {\bf{k}}_F , i0 , \sigma  )
 \nonumber
 \\
  & & \hspace{-72mm} \equiv  \Gamma_0^{(4)} 
(  {\bf{k}}_F, \sigma   ; {\bf{k}}_F^{\prime} , \sigma^{\prime} )
   \; ,
 \label{eq:Landaudef} 
\end{eqnarray}
i.e. we project all momenta onto the Fermi surface, ignore the
frequency dependence, and replace the flowing vertex by its
fixed point value.
Moreover, at this level of approximation we may also replace
$ \Sigma_{ \xi_{ {\bf{k}}^{\prime}   } } ( K^{\prime} ) 
 \rightarrow \Sigma ( {\bf{k}}_F^{\prime} , i0 )$ on the right-hand side of 
Eq. (\ref{eq:counterint}).  For $\xi_0 \rightarrow \infty$ we then obtain
the Hartree-Fock self-consistency equation
 \begin{eqnarray}
 \Sigma ( {\bf{k}}_F , i0 )  & = &  \sum_{\sigma^{\prime} } 
 \int \frac{ d {\bf{k}}^{\prime}   }{ (2 \pi )^{D}} 
 \Gamma_0^{(4)} 
(  {\bf{k}}_F, \sigma   ; {\bf{k}}_F^{\prime} , \sigma^{\prime} )
 \nonumber
 \\
 & \times &
 \Theta \left( \mu - 
 \epsilon_{ {\bf{k}}^{\prime}}  - \Sigma ( {\bf{k}}_F^{\prime} , i0 ) \right)
 \; .
 \label{eq:MF}
 \end{eqnarray}
For a given interaction vertex
$  \Gamma_0^{(4)} 
(  {\bf{k}}_F, \sigma   ; {\bf{k}}_F^{\prime} , \sigma^{\prime} )   $,  
Eq. (\ref{eq:MF}) 
can be used to study possible
Fermi surface instabilities such as the Pomeranchuk instability \cite{Pomenanchuk58} at
the Hartree-Fock level, which should always be
the first step before more elaborate methods are used.
A trivial generalization of Eq. (\ref{eq:MF}) with a 
spin-dependent  counterterm  $\Sigma ( {\bf{k}}_F , i0 , \sigma )$ 
leads to the self-consistent Hartree-Fock equation for spontaneous
ferromagnetism, implying the usual  Stoner instability
for strong enough interactions in the spin-channel.
More generally, if we allow for other types of symmetry
breaking, within the same approximations as above 
the exact RG fixed point equation for the
two-point vertex can be reduced to
the  Hartree-Fock self-consistency equation for the corresponding 
order parameter.
Note  that no approximation has been made in deriving  
Eqs. (\ref{eq:integral2}) and (\ref{eq:counterint}), so that 
these exact fixed point equations (or generalizations thereof for 
other types of symmetry breaking)
can serve as  a starting point
for a systematic calculation of corrections to the Hartree-Fock approximation.

In summary, in this work we have shown how the true Fermi surface of 
an interacting Fermi system  can be defined self-consistently as a fixed point
property of the RG. Our main result are the two equivalent 
integral equations (\ref{eq:integral2}) and (\ref{eq:counterint}), 
which determine the counterterm $\Sigma ( {\bf{k}}_F , i0 )$
necessary to calculate the Fermi surface from Eq. (\ref{eq:FSdef}). 
In the simplest approximation, Eq. (\ref{eq:counterint}) reduces 
to the Hartree-Fock self-consistency
condition for the counterterm. However, systematic improvements are possible.

Very recently  Dusuel and Dou\c{c}ot \cite{Dusuel02}
presented a detailed analysis of the Fermi surface deformations in
quasi one-dimensional electronic systems, using perturbation theory and
the RG method. 
They realized that some ``slight modification'' of the Wilson-Polchinski RG
approach is necessary in order to use this approach for a self-consistent
calculation of the Fermi surface,  but admitted that
a practical implementation of such a modification remains to be attempted.
We have shown here that 
the necessary modification of the functional RG 
used in Refs. \cite{Zanchi96,Salmhofer98,Halboth00,Binz02}
is simply the usual rescaling step \cite{Kopietz01},
which is an essential part of the
``orthodox'' Wilsonian RG \cite{Ma76}. 
We conclude that
the functional RG approach for interacting Fermi systems 
in the form advanced  in Ref. \cite{Kopietz01} 
provides an elegant solution to the difficult \cite{Nozieres64} problem
of self-consistently constructing the true Fermi surface.

We thank Lorenz Bartosch, Tom Busche,  Alvaro Ferraz, Volker Meden, Walter Metzner 
for discussions.
This work was partially supported by the DFG via 
Forschergruppe FOR 412, Project No. KO 1442/5-1.


\begin{thebibliography}{99}
\bibitem{footnotereal}
We assume that $\Sigma ( {\bf{k}}_F , i0 )$ is real, which is always true 
for Fermi liquids, where the damping of quasiparticles with wavevectors 
precisely on the Fermi surface vanishes.
\bibitem{footnotedens}
To compare the Fermi surface of the interacting system
with the corresponding Fermi surface without interactions at the same density $n$, 
we should first calculate the density $n ( \mu )$ 
corresponding to the Fermi surface ${\bf{k}}_F$ of the interacting system
from Eq. (\ref{eq:FSdens}), and then determine
$\mu_0$ such that
$n ( \mu ) =   \sum_{\sigma } 
 \int \frac{ d {\bf{k}}   }{ (2 \pi )^{D}} \Theta (  \mu_0 -    \epsilon_{ {\bf{k}}}  )$.
Given $\mu_0$,   we may calculate the Fermi surface of the non-interacting system 
at the density $n ( \mu )$ from
the roots of  $ \epsilon_{ {\bf{k}}  }= \mu_0 $.
\end{thebibliography}
\end{document}